\documentclass[aps,prb,reprint,twocolumn,superscriptaddress,amsmath,amssymb,10pt]{revtex4-2}
\usepackage{graphicx}
\usepackage{epstopdf}
\usepackage{xcolor}
\usepackage[colorlinks, linkcolor=red,citecolor=blue,urlcolor=blue]{hyperref}
\usepackage[all]{hypcap} 
\begin{document}
\bibliographystyle{apsrev}

\title{Disorder-induced trapping and anti-trapping of vortices in type-II superconductors}

\author{A.~A.~Kopasov}
\affiliation{Institute for Physics of Microstructures, Russian Academy of Sciences, 603950 Nizhny Novgorod, GSP-105, Russia}
\affiliation{Lobachevsky State University of Nizhny Novgorod, 603950 Nizhny Novgorod, Russia}
\author{I.~M.~Tsar'kov}
\affiliation{Lobachevsky State University of Nizhny Novgorod, 603950 Nizhny Novgorod, Russia}
\author{A.~S.~Mel'nikov}
\affiliation{Institute for Physics of Microstructures, Russian Academy of Sciences, 603950 Nizhny Novgorod, GSP-105, Russia}
\affiliation{Lobachevsky State University of Nizhny Novgorod, 603950 Nizhny Novgorod, Russia}

\date{\today}

\begin{abstract}
We study the features of the superconductivity nucleation and vortex configurations in superconductors with modulated disorder. Using the Ginzburg-Landau-type theory with spatially varying diffusion coefficient, we uncover and explain the switching between the vortex-defect attraction to the repulsion upon the increase in the external magnetic field. It is shown that for rather weak applied magnetic fields, a superconducting nucleus localized near the region with the suppressed diffusion coefficient possesses a nonzero vorticity whereas the increase in the magnetic field can lead to a transition into the state with zero winding number. We demonstrate the manifestations of this switching phenomenon in superconductors with a large number of defects by performing numerical simulations of the vortex structures in superconductors with periodic spatial profiles of the diffusion coefficient. The obtained results clarify the physics of the vortex arrangement in several classes of the superconducting materials including one-dimensional superlattices and nanopatterned superconductors with regular arrays of the defects characterized by the increased concentration of nonmagnetic impurities.
\end{abstract}

\maketitle

\section{Introduction}\label{introduction}

The physics of the vortex pinning mechanisms in type-II superconductors is known to be a fundamental problem, which is also of particular importance for a variety of applications of superconducting materials~\cite{Campbell1972,Blatter1994}. Therefore, for more than half a century the effects of vortex pinning have been in the focus of both theoretical and experimental works studying the vortex interaction with various types of inhomogeneities including columnar defects~\cite{BezryadinPLA1994,BezryadinJLTP1995,BerdiyorovEPL2006,BerdiyorovPRL2006,BerdiyorovPRB2006,SabatinoJAP2010,LatimerPRB2012,LatimerPRL2013,GePRB2017,XueNJP2018}, blind holes~\cite{BezryadinPRB1996,BerdiyorovNJP2009}, non-superconducting inclusions~\cite{KarapetrovPRL2005,SadovskyyPRB2017}, etc. General mechanisms of the vortex - defect interaction are associated either with the defect-induced change of the energy of supercurrents flowing around the vortex axis~\cite{BeanPRL1971} or with the changes of the vortex core energy~\cite{Campbell1972,Blatter1994}. The development of technology provided a number of experimental methods widely used to improve the vortex pinning characteristics including, e.g., the ion irradiation techniques~\cite{BugoslavskyN2001,NakajimaPRB2009,ZechnerSUST2018,AntonovPSS2019,AntonovPSS2020,AntonovPC2020,AichnerFNT2020}. It is interesting to note that the pinning centers can appear even without strong local change in the material properties resulting in variations of the superconducting critical temperature $T_c$ or creation of insulating inclusions. The vortex core energy can be effectively changed due to spatial modulation of quasiparticle scattering characteristics~\cite{ThunebergPRB1984}. Indeed, for superconductors in the dirty limit the size of the vortex core $r_c$ is given by the superconducting coherence length $\xi_0 \propto \sqrt{D/T_c}$, where $D = v_F\ell/3$ is the diffusion coefficient, $v_F$ is the Fermi velocity, and $\ell$ is the mean free path for impurity scattering. The increase in the local impurity concentration leads to the  suppression of the mean free path $\ell$ and the size of the vortex core, which, in turn, determines the energy per unit length of the vortex line $\epsilon = \epsilon_c r_c^2$, where $\epsilon_c$ is the superconducting condensation energy density. Thus, according to general belief, the regions with the increased disorder attract Abrikosov vortices.

\begin{figure}[htpb]
\centering
\includegraphics[scale = 0.65]{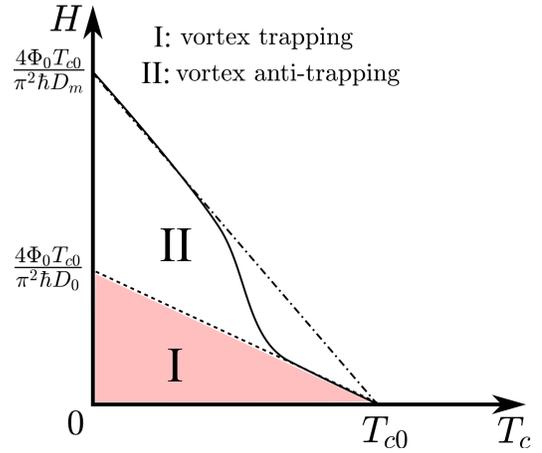}
\caption{Schematic superconducting phase diagram magnetic field $H$ - temperature $T$ for disordered superconductor with the diffusion coefficient $D_0$ containing a mesoscopic region with suppressed diffusion coefficient $D_m<D_0$. Solid (dotted) line shows typical $T_c(H)$ curve for a superconducting nucleus localized at the defect (far away from it). Dashed-dotted line shows $T_c(H)$ curve for the bulk superconductor with the diffusion constant $D_m$. Within the region I (II) of the phase diagram the defect attracts (repels) Abrikosov vortices. Here $\Phi_0 = \pi\hbar c/|e|$ is the magnetic flux quantum.}
\label{Fig:Fig1}
\end{figure}

The main goal of the present manuscript is to show that the validity range of the above arguments is restricted to the case of weak magnetic fields and to predict the phenomenon of switching from the vortex-defect attraction to the repulsion at sufficiently strong magnetic fields. The appearance of such drastic change in the flux pinning mechanism can occur when the magnetic length $L_H = \sqrt{\Phi_0/2\pi H}$ becomes comparable to a typical size of the regions with increased disorder. Here $\Phi_0 = \pi\hbar c/|e|$ is the magnetic flux quantum and $H$ is the applied magnetic field. In order to clarify this statement, let us, first, recall the basic features of the superconductivity nucleation at an isolated defect characterized by the diffusion constant $D_m$, which is embedded into the superconductor with larger diffusion coefficient $D_0>D_m$. For definiteness, we consider a cylindrical defect with the cross-section radius $R\sim\xi_0\propto\sqrt{D_0/T_c}$, and the system is subjected to the external magnetic field aligned with the cylinder axis. It is well-established that mesoscopic regions with enhanced quasiparticle scattering reveal themselves in a change of the shape of superconducting phase transition curve magnetic field $H$ - temperature $T$ for superconducting nucleus localized at the defect~\cite{TakahashiPRB11986,TakahashiPRB21986,KopasovRQE2017,KopasovPRB2017} (see the solid line in Fig.~\ref{Fig:Fig1}). The resulting enhancement of the superconducting critical temperature occurs at sufficiently strong magnetic fields $H \gtrsim \Phi_0/2\pi R^2$, for which the size of the superconducting nucleus is determined by the local value of the diffusion coefficient $D_m$.  Thus, the superconducting $H$-$T$ phase diagram can be divided into two regions (schematically shown as regions I and II in Fig.~\ref{Fig:Fig1}) characterized by different spatial distribution of the Cooper pair density (superconducting condensation energy density). For rather weak magnetic fields (region I), the superconductivity is developed in the whole sample. In this case the above mentioned core pinning argument is correct, so that the regions with the increased disorder trap vortices. The situation is qualitatively different in the opposite case of strong magnetic fields (region II). In this regime the superconductivity is mainly developed at the defect whereas it is exponentially suppressed outside. Therefore, despite the fact that the vortex core size can be larger outside the defect, it is energetically more favorable for vortices to escape from the defect and to be pinned by the regions with significantly reduced superconducting condensation energy density $\epsilon_c$.

\begin{figure}[htpb]
\centering
\includegraphics[scale = 0.95]{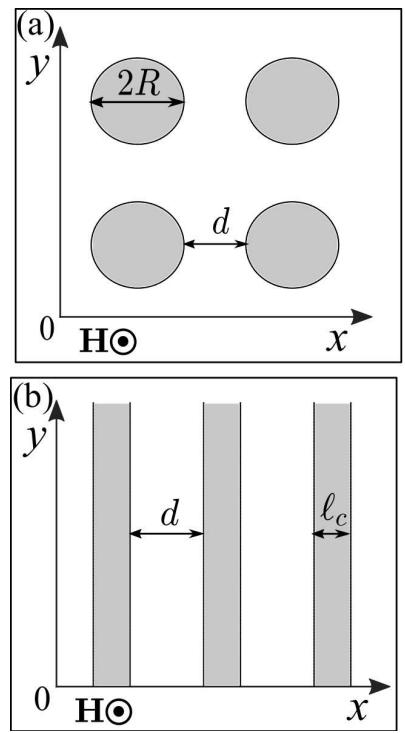}
\caption{Schematic of spatially periodic profiles of the diffusion coefficient $D(\mathbf{r})$ considered in the present work. In the white (gray shaded) regions $D(\mathbf{r}) = D_0$ ($D_m$) and $D_0/D_m > 1$. Panel (a) shows a regular array of the defects with a circular cross section defined by the defect radius $R$ and the inter-defect distance $d$. Panel (b) shows the one-dimensional superlattice characterized by the layer thickness $\ell_c$ and the interlayer distance $d$. For both types of systems we choose the external magnetic field $\mathbf{H}$ to be directed along the $z$-axis.}
\label{Fig:diffusion_distribution}
\end{figure}

To provide a quantitative consideration of the above switching in the pinning mechanism, we study the features of the superconductivity nucleation and the vortex phases in superconductors with modulated disorder. Based on the Ginzburg-Landau (GL) theory with spatially varying diffusion coefficient, we show that a superconducting nucleus localized near the region with the suppressed diffusion coefficient can possess a nonzero vorticity. Corresponding analysis mainly relies on the calculations of the $H$-$T$ phase-transition lines for several types of isolated inhomogeneities, which include a cylindrically-shaped region in the bulk material or a disc in a thin superconducting film as well as a layer of a finite thickness. We find a sudden change in the spatial structure of the emerging superconducting order parameter at a certain point at the phase-transition line $(H^+,T^+)$. For $H < H^+$ ($T>T^+$) a superconducting nucleus possesses a nonzero winding number while the superconducting state with zero vorticity appears for $H > H^+$ ($T<T^+$). Such peculiarities of individual vortex trapping and anti-trapping at a single defect area should reveal themselves in rearrangements of the vortex arrays in macroscopic samples. To avoid a rather complicated analysis of the vortex matter in samples with random distributions of defect regions~\cite{LarkinJETP1970,LarkinJETP1972}, we restrict ourselves to the illustration of the switching phenomena with regular spatial distributions of the impurity concentration. For this purpose, we perform direct numerical simulations of the vortex phases in several types of superconducting superstructures by solving the nonlinear GL equation. The effects of the modulated disorder on the vortex arrangement have been analyzed for two types of systems which include bulk superconductors with an embedded square array of cylindrical defects or thin superconducting films with a square array of discs (see Fig.~\ref{Fig:diffusion_distribution}(a)) as well as the one-dimensional superconducting superlattices (see Fig.~\ref{Fig:diffusion_distribution}(b)). It is remarkable that the resulting vortex arrangements in these two geometrically different types of superstructures possess some similarities. It is shown that for sufficiently weak external magnetic fields, all the vortices are pinned by the regions with the suppressed diffusion coefficient. The opposite is true in the case of rather strong magnetic fields, namely the vortices are located only between inhomogeneities. Within the intermediate field range, the vortex structure consists of two types of vortices with different core sizes located near the regions with the increased impurity concentration and in between them.

It is important to note that the validity of our results is restricted to the case of weak disorder, so that the superconducting critical temperature at zero magnetic field is homogeneous throughout the sample~\cite{Anderson1959,Abrikosov1959}. This regime can take place in a variety of systems including superconductors irradiated with sufficiently low doses of ions and hybrid systems composed of superconducting materials with nearly the same critical temperatures and different disorder characteristics~\cite{KarkutPRL1988,KuwasawaPC1990,AartsPB1990,AartsEPL1990,KuwasawaPC1991,NojimaPC1993,KoorevaarPRB1993,NojimaPB1994,KuwasawaPB1996,ObiPSS2001}. Note also that calculations of the vortex arrangement performed in our work rely neither on the assumption of a weak spatial modulation of the diffusion coefficient~\cite{AmiPTP1975} nor on specific assumptions about the structure of the vortex lattice~\cite{AmiPTP1975,YuanZP1995}. Direct numerical solution of the nonlinear GL equation allows to capture possible transitions between different vortex configurations and to confirm our qualitative arguments regarding the switching between disorder-induced trapping and anti-trapping of vortices upon the increase in the applied magnetic field.

This manuscript is organized as follows. In Sec.~\ref{model_methods} we present the model and briefly describe the 
methods. In Sec.~\ref{isolated_droplets} we carry out the detailed analysis of the phase-transition lines magnetic field - temperature for several types of isolated inhomogeneities. In Sec.~\ref{vortex_phases} we present the results of numerical simulations of the vortex phases in superconductors with spatially periodic distribution of the impurity concentration. Finally, the results are summarized in Sec.~\ref{discussion}.

\section{Model and methods}\label{model_methods}

Hereafter, we consider conventional superconductors with spatially varying concentration of nonmagnetic impurities described by an inhomogeneous diffusion coefficient. Our theoretical approach involves two basic assumptions. First, we assume that the disorder is too weak to affect the local density of states and the BCS coupling constant, so that the conditions of the Anderson theorem~\cite{Anderson1959,Abrikosov1959} are fulfilled. Second, we consider the case when the magnetic field penetration length greatly exceeds the spatial scale of the superconducting correlations, which allows us to neglect the effects of the Meissner screening. Corresponding GL free energy $F = \int f(\mathbf{r})d^3\mathbf{r}$ reads
\begin{equation}\label{GL_free_energy}
 f(\mathbf{r}) = a|\Delta|^2 + \gamma(\mathbf{r}) |\hat{\mathbf{\Pi}}\Delta|^2 + \frac{b}{2}|\Delta|^4  \ .
\end{equation}
Here $\Delta(\mathbf{r}) = |\Delta(\mathbf{r})|\exp[i\chi(\mathbf{r})]$ is the superconducting order parameter, $a = -\alpha(T_{c0}-T)$, $b$, and $\gamma(\mathbf{r})$ are the GL coefficients, $\hat{\mathbf{\Pi}} = (i\nabla - 2\pi\mathbf{A}/\Phi_0)$, and $\mathbf{A}$ is the vector potential. Minimization of $F$ with respect to $\Delta^*$ gives the GL-type equation
\begin{equation}\label{nonlinear_GL_equation}
 \hat{\mathbf{\Pi}}\zeta^2(\mathbf{r})\hat{\mathbf{\Pi}}\Psi - \Psi + |\Psi|^2\Psi = 0 \ ,
\end{equation}
where $\Psi(\mathbf{r}) = \Delta(\mathbf{r})/\Delta_{\infty}$, $\Delta_{\infty} = \sqrt{|a|/b}$, and $\zeta(\mathbf{r})$ is the superconducting coherence length  
\begin{equation}\label{temperature_dependent_coherence_length}
 \zeta(\mathbf{r}) = \sqrt{\frac{\gamma(\mathbf{r})}{|a|}} = \sqrt{\frac{\pi\hbar D(\mathbf{r})}{8(T_{c0} - T)}} \ .
\end{equation}

In the following section we analyze the features of the superconductivity nucleation at single isolated inhomogeneities described by the following $D(\mathbf{r})$ profiles:
\begin{subequations}\label{diffusion_profiles}
 \begin{align}
  \label{2D_diffusion_profile}
  D(\rho) = D_0 + \delta D\Theta(R - \rho) \ ,\\
  \label{1D_diffusion_profile}
  D(x) = D_0 + \delta D\left[\Theta(x + \ell_c/2) - \Theta(x-\ell_c/2)\right] \ .
 \end{align}
\end{subequations}
Here $\rho$ is the radial coordinate in a plane perpendicular to the direction of the applied magnetic field, $\delta D = -(D_0 - D_m)$, $D_0/D_m \geq 1$, and $\Theta(x)$ is the Heaviside function. Corresponding analysis mainly relies on calculations of the superconducting phase-transition lines magnetic field - temperature which are determined from the lowest eigenvalue of the problem
\begin{equation}\label{GL_linearized}
 \boldsymbol{\Pi}\xi^2(\mathbf{r})\boldsymbol{\Pi}\Psi = E_0\Psi \ ,
\end{equation}
where $E_0(H) = 1 - T_c(H)/T_{c0}$, $T_c$ is the superconducting critical temperature, and $\xi(\mathbf{r})$ is the zero-temperature superconducting coherence length defined by Eq.~(\ref{temperature_dependent_coherence_length}) for $T = 0$. Numerical solution of the eigenvalue problem~(\ref{GL_linearized}) has been carried out by matching the exact solutions of the linearized GL equation in the regions with a constant diffusion coefficient (see Appendix~\ref{appendix} for the details).

\begin{figure*}[htpb]
\centering
\includegraphics[scale = 0.9]{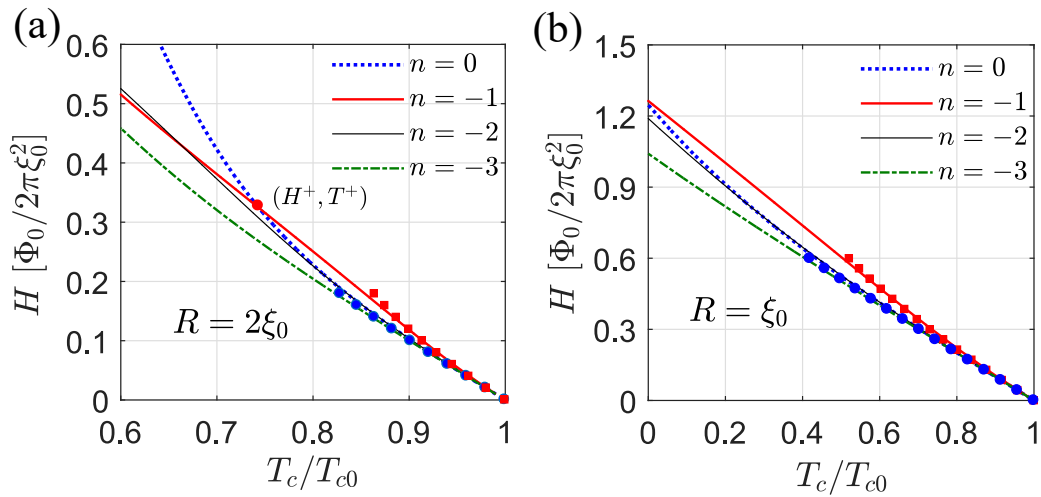}
\caption{Phase-transition lines $T_c(H)$ for the superconducting droplets with vorticities $n = 0,-1,-2,-3$ for the two-dimensional $D(\mathbf{r})$ profile~(\ref{2D_diffusion_profile}). Panels (a) and (b) correspond to $R = 2\xi_0$ and $R = \xi_0$, respectively, and $\xi_0 = \sqrt{\pi \hbar D_0/8T_{c0}}$. We take $D_0/D_m = 3$ to produce the plots. Lines show the numerical results whereas filled circles and squares correspond to the results of the perturbation theory~(\ref{2D_perturbation_theory}).}
\label{Fig:2D}
\end{figure*}

The analysis of the vortex configurations in superconducting superstructures in Sec.~\ref{vortex_phases} is based on numerical solutions of the nonlinear GL equation~(\ref{nonlinear_GL_equation}). For this purpose we employ the relaxation method with an addition of the time derivative $\partial\Psi/\partial t$ into the right-hand side of Eq.~(\ref{nonlinear_GL_equation}) and look for the solution $\Psi(x,y)$, which doesn't depend on time. We use the implicit Crank-Nicolson integration scheme with a linearization of the nonlinear $|\Psi|^2\Psi$-term for the time integration~\cite{SadovskiiJCP2015}. The link variable approach~\cite{KatoPRB1993} is used for the spatial discretization of the GL equation on a square simulation area $x, y\in[0, L]$, where $L$ is an integer multiple of the superstructure period. In numerical simulations we choose the gauge $\mathbf{A} = H(x - L_x/2)\hat{\mathbf{y}}$, and the magnitude of the external magnetic field $H$ is parametrized by an integer number $n_q$ of the magnetic flux quanta $\Phi_0$ piercing through the simulation area $HL^2 = n_q\Phi_0$. The periodic boundary conditions for the Cooper-pair wave function, which simulate the periodicity of the spatial profiles of the diffusion coefficient, have the following form~\cite{BerdiyorovPRB2006}: 
\begin{subequations}\label{quasiperiodic_boundary_conditions}
 \begin{align}
  \Psi(L, y) = \exp\left(-2\pi i \eta_x\right)\Psi(0,y) \ ,\\
  \Psi(x,L) = \exp\left(-2\pi i \eta_y \right)\Psi(x,0) \ .
 \end{align}
\end{subequations}
Here $\eta_x(y) = HLy/\Phi_0 + C_x$, $\eta_y = C_y$, and $C_{x,y}$ are real constants, which can be determined from the condition that the superfluid velocity $\mathbf{v}_s\propto [\mathbf{A} + (\Phi_0/2\pi)\nabla\chi]$ averaged over the simulation area is zero. For our choice of the gauge and the simulation region this condition fixes $C_x = -n_q/2$ and $C_y = 0$.

\section{Phase-transition lines for isolated superconducting droplets}\label{isolated_droplets}

In this section we demonstrate the possibility of switching in the pinning mechanism for isolated regions with increased disorder. In particular, we consider the phase-transition lines and the spatial structure of the emerging superconducting order parameter for the two- and one-dimensional profiles of the diffusion coefficient~(\ref{2D_diffusion_profile}) and (\ref{1D_diffusion_profile}).

We start our study from the two-dimensional case and take an exemplary $D(\mathbf{r})$ profile~(\ref{2D_diffusion_profile}). Choosing the radial gauge $\mathbf{A} = H\rho\boldsymbol{\varphi}_0/2$, one finds the solution of Eq.~(\ref{GL_linearized}) as $\Psi(\mathbf{r}) = \psi_{n,k_z}(\rho)\exp\left(in\varphi + ik_zz\right)$, and the minimal eigenvalue of the problem~(\ref{GL_linearized}) corresponds to $k_z = 0$. The radial part of the superconducting order parameter $\psi_{n,k_z = 0}(\rho)$ can be expressed via the generalized Laguerre polynomial (see Appendix~\ref{appendix}). Treating the spatial modulation of the diffusion coefficient within the first-order perturbation theory, we obtain the phase transition lines $T_c(H)$ for superconducting droplets with vorticities $n = 0$ and $-1$
\begin{subequations}\label{2D_perturbation_theory}
 \begin{align}
 \label{zero_vorticity}
 \frac{T_c^{n =0}(H)}{T_{c0}}\approx 1 - \frac{\xi_0^2}{L_H^2}\left[1 + \frac{1}{8}\frac{\delta D}{D_0}\left(\frac{R}{L_H}\right)^4\right] \ ,\\
 \label{unit_vorticity}
 \frac{T_c^{n = -1}(H)}{T_{c0}}\approx 1 - \frac{\xi_0^2}{L_H^2}\left[1 + \frac{1}{2}\frac{\delta D}{D_0}\left(\frac{R}{L_H}\right)^2\right] \ .
 \end{align}
\end{subequations}
Here $\xi_0 = \sqrt{\pi\hbar D_0/8T_{c0}}$ and $L_H = \sqrt{\Phi_0/2\pi H}$ is the magnetic length. Note that the validity of Eqs.~(\ref{zero_vorticity}) and (\ref{unit_vorticity}) is also restricted to the limit $R/L_H\ll 1$. The above expressions indicate that for rather weak magnetic fields, superconductivity nucleates in the form of a vortex pinned by the region with increased impurity concentration. The results of numerical calculations of $T_c(H)$ curves for droplets with vorticities $n = 0$, $-1$, $-2$, and $-3$ are shown in Fig.~\ref{Fig:2D}. Panels (a) and (b) correspond to $R = 2\xi_0$ and $R = \xi_0$, respectively. We take $D_0/D_m = 3$ to produce the plots. One can see that the results in Fig.~\ref{Fig:2D}(a) reveal a sudden change in the spatial structure of the emerging superconducting order parameter at a certain point at the phase-transition line $(H^+,T^+)$. For $H<H^+$ ($T>T^+$) the superconducting nucleus has the form of a singly quantized vortex pinned by the region with the suppressed diffusion coefficient whereas the localized superconducting state with zero vorticity appears for $H>H^+$ ($T<T^+$). Thus, at rather strong magnetic fields the vortex escapes from the defect region. Comparing the results in Figs.~\ref{Fig:2D}(a) and~\ref{Fig:2D}(b), one can see that the position of the intersection point ($H^+,T^+$) depends on the size of a region with the suppressed diffusion coefficient. Rough estimates for $H^+$ and $T^+$ can be obtained from the condition that the defect radius is comparable to the characteristic size of the superconducting droplet 
\begin{subequations}\label{estimates}
 \begin{align}
 H^+ \sim \Phi_0/R^2 \ ,\\
 T^+ \sim T_{c0}\left[1 - (\xi_0/R)^2\right] \ .
 \end{align}
\end{subequations}
Correspondingly, for rather small $R$ values, the superconducting nucleus should possess a nonzero vorticity within the full temperature and magnetic field range which is in agreement with the results shown in Fig.~\ref{Fig:2D}(b).

\begin{figure*}[htpb]
\centering
\includegraphics[scale = 0.9]{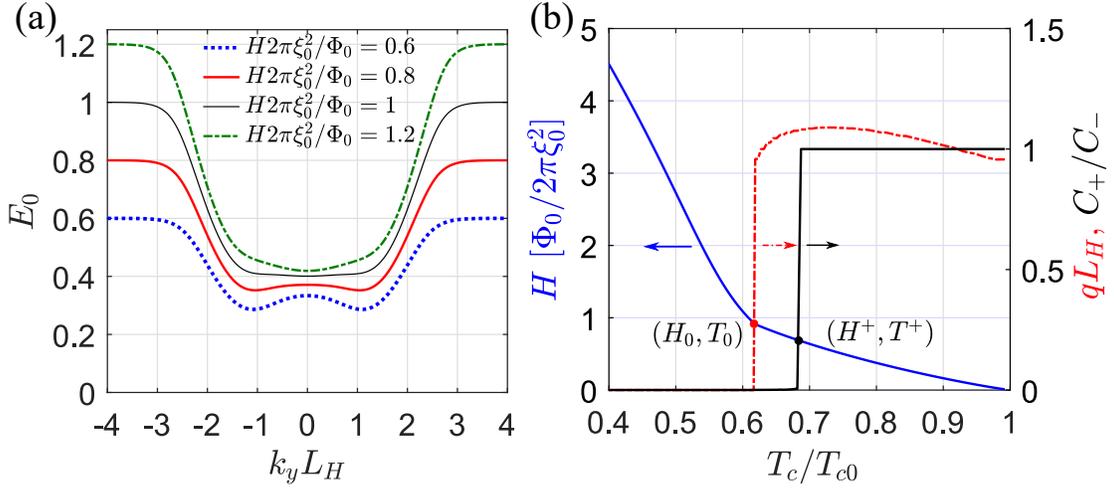}
\caption{(a) Minimal eigenvalue $E_0$ of the problem~(\ref{GL_linearized}) for the one-dimensional $D(\mathbf{r})$ profile~(\ref{1D_diffusion_profile}) versus $k_y$ for $H/(\Phi_0/2\pi\xi_0^2) = 0.6$, 0.8, 1, 1.2. Panel (b) shows the resulting phase-transition curve $T_c(H)$ (blue solid line), $k_y = q$ values maximizing the critical temperature (red dashed-dotted line), and the temperature dependence of the ratio $C_+/C_-$ corresponding to the minimum of the Abrikosov parameter~(\ref{Abrikosov_parameter}) for a trial function~(\ref{1D_trial_function}). Only the branch with positive $q$ is shown. We take $D_0/D_m = 10$ and $\ell_c = 1.4\xi_0$ to produce the plots.}
\label{Fig:1D}
\end{figure*}

Let us proceed with the analysis of the one-dimensional $D(\mathbf{r})$ profile~(\ref{1D_diffusion_profile}). Choosing the Landau gauge $\mathbf{A} = (0,Hx,0)$, the solution of Eq.~(\ref{GL_linearized}) can be presented in the form $\Psi(\mathbf{r}) = \psi_{k_y,k_z}(x)\exp\left[i\left(k_yy + k_zz\right)\right]$, and the minimal eigenvalue of the problem~(\ref{GL_linearized}) corresponds to $k_z = 0$. The function $\psi_{k_y,k_z = 0}(x)$ can be expressed via the Hermite function (see Appendix~\ref{appendix}). Within the first-order perturbation theory over the spatial modulation of the diffusion coefficient and in the limit $\ell_c/L_H \ll 1$, the dispersion of the lowest Landau level is determined by the function
\begin{equation}\label{1D_perturbation_theory}
 E_0(k_y) \approx \frac{\xi_0^2}{L_H^2} - \frac{2(\xi_0^2 - \xi_m^2)}{\sqrt{\pi}L_H^2}\frac{\ell_c}{L_H}(k_yL_H)^2e^{-(k_yL_H)^2} \ ,
\end{equation}
which possesses two degenerate minima at $k_y = \pm 1/L_H$. Here $\xi_m = \sqrt{\pi\hbar D_m/8T_{c0}}$. Correspondingly, the spatial modulation of the diffusion coefficient stabilizes superconducting droplets, which are centered at $x = \mp L_H$. This picture changes qualitatively upon the increase in the external magnetic field. To go beyond the validity range of Eq.~(\ref{1D_perturbation_theory}), we carry out numerical calculations of the phase-transition lines 
\begin{equation}
 T_c/T_{c0} = 1 - \min_{k_y}\left[E_0(k_y)\right] \ .
\end{equation}
We find the lowest eigenvalue $E_0(k_y)$ of the problem~(\ref{GL_linearized}) and also extract the corresponding $k_y = q$ values, which minimize $E_0$ (maximize the critical temperature $T_c$). Typical $E_0(k_y)$ curves for $H/(\Phi_0/2\pi\xi_0^2) = 0.6$, 0.8, 1, and 1.2 are presented in Fig.~\ref{Fig:1D}(a). We take $D_0/D_m = 10$ and $\ell_c = 1.4\xi_0$ to produce the plots. One can see that the increase in the applied magnetic field leads to a qualitative change in the nature of the minimization for $E_0$ with respect to $k_y$. At a certain value of the magnetic field $H_0$, the positions of the minima on $E_0(k_y)$ curves exhibit a jumpwise change from finite values to zero, which corresponds to a jumpwise change in the position of the superconducting nucleus. For $H>H_0$ ($H<H_0$) the droplet is centered at $x \approx \pm L_H$ ($x = 0$). Note that rough estimates for $H_0$ and $T_0$ are given by Eqs.~(\ref{estimates}) with the replacement $R\to\ell_c$. The resulting phase-transition curve $T_c(H)$ along with the corresponding $q(T_c)$ dependence are shown in Fig.~\ref{Fig:1D}(b). The obtained results indicate that a jumpwise change on $q(T_c)$ curve reveals itself through a change in the slope on the $T_c(H)$ curve. 


In the vicinity of the phase-transition line, one can obtain the spatial structure of the superconducting order parameter by minimizing the GL free energy~(\ref{GL_free_energy}) for a trial function~\cite{SJ}
\begin{equation}\label{1D_trial_function}
 \Psi(\mathbf{r}) = C_+\psi_{k_y}(x)e^{ik_yy} + C_-\psi_{k_y}(-x)e^{-ik_yy} \ ,
\end{equation}
where the constants $C_{\pm}$ can be chosen to be real numbers. This procedure is equivalent to the minimization of the Abrikosov parameter
\begin{equation}\label{Abrikosov_parameter}
 \beta_A = \frac{\int |\Psi(\mathbf{r})|^4d^3\mathbf{r}}{\left(\int |\Psi(\mathbf{r})|^2d^3\mathbf{r}\right)^2} \ .
\end{equation}
As a next step, we substitute $k_y = q(T_c)$ and $\psi_{k_y}(x) = \psi_{q}(x)$ from numerical calculations into the above equation and perform the minimization of $\beta_A$ with respect to the ratio $C_{+}/C_{-}$ (or equivalently $C_-/C_+$). The results of such procedure are shown in Fig.~\ref{Fig:1D}(b), in which we plot $C_+/C_-$ values corresponding to the minimum free energy in the vicinity of the superconducting phase transition. Thus, the spatial distribution of the Cooper-pair wave function can be characterized as follows. For $H < H^+$ ($T>T^+$) we get $C_{+} = C_{-}$. In this regime, the layer with the suppressed diffusion coefficient hosts a chain of Abrikosov vortices centered at $x = 0$. In the opposite case $H > H^+$ ($T<T^+$) the vortices escape from the regions with increased disorder. Indeed, within the intermediate magnetic field and temperature range $H_0<H<H^+$ ($T_0<T<T^+$) one gets a doubly degenerate state characterized by the following sets of coefficients in Eq.~(\ref{1D_trial_function}): $C_+\neq 0$, $C_- = 0$ and $C_+ = 0$, $C_-\neq 0$. In other words, the superconducting dropet is shifted either to the left or to the right from the center of the layer with the suppressed diffusion coefficient. Finally, in the strong-field (low-temperature) regime $H>H_0$ ($T<T_0$), the maximum modulus of the superconducting order parameter is at the center of the layer with the increased impurity concentration. Note that the relative positions of the points $(H^+,T^+)$ and $(H_0,T_0)$ at the phase-transition line are governed by the ratio of the diffusion constants $D_0/D_m$ and the thickness of the layer with the suppressed diffusion coefficient $\ell_c$. In numerical simulations we observe that for rather large $D_0/D_m$ ratios and small $\ell_c$ values $H^+ < H_0$ ($T^+ > T_0$) whereas in the opposite case these points coincide $H^+ = H_0$ ($T^+ = T_0$).

\section{Vortex phase transitions in superconducting superstructures}\label{vortex_phases}

\subsection{Square array of cylindrical defects}

We proceed with a discussion of the results of numerical simulations of the vortex states. Here we consider the bulk superconductors containing containing square array of cylindrical defects with the suppressed diffusion constant (see Fig.~\ref{Fig:diffusion_distribution}(a)). The obtained results are also relevant for thin superconducting films with an embedded array of discs provided that the film thickness is much smaller than the superconducting coherence length and the magnetic field penetration depth.

\begin{figure}[htpb]
\centering
\includegraphics[scale = 1.1]{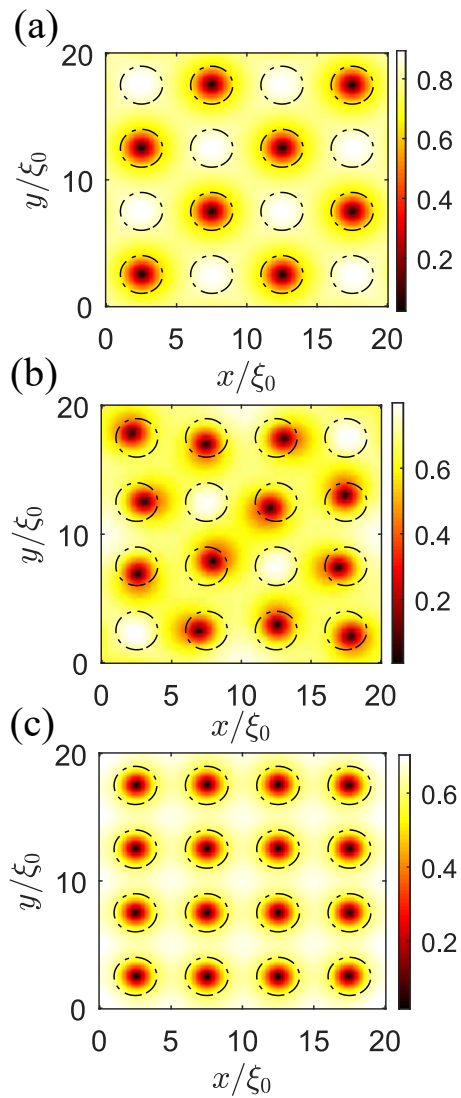}
\caption{Color plots of the absolute value of the Cooper-pair wave function $|\Psi|$ for periodic $D(\mathbf{r})$ profile shown in Fig.~\ref{Fig:diffusion_distribution}(a). Panels (a), (b), and (c) correspond to $n_q = 8$, $12$, and $16$, respectively, where $n_q$ is the number of the magnetic flux quanta in the computational cell. Black dashed-dotted lines highlight the boundaries between the regions with different diffusion coefficient. The parameters are: $T = 0.8T_{c0}$, $D_0/D_m = 3$, $R = 1.5\xi_0$, and $d = 2\xi_0$.
}
\label{Fig:cylinders_low_fields}
\end{figure}

Let us, first, discuss the regime of rather weak external magnetic fields $|H| \leq H^*$, where $H^*$ is defined from the condition that the total magnetic flux piercing through the simulation area is equal to the number of defects (the so-called first matching field)
\begin{equation}
 H^* = \Phi_0/(2R + d)^2 \ .
\end{equation}
The resulting color plots of the absolute value of the Cooper-pair wave function $|\Psi(x,y)|$ for several numbers of the magnetic flux quanta in the computational cell $n_q = 8$, $12$, and $16$ are shown in Fig.~(\ref{Fig:cylinders_low_fields}). One can see that for $n_q = 8$ (Fig.~\ref{Fig:cylinders_low_fields}(a)), the vortices form a square lattice with twice the periodicity of the diffusion coefficient distribution. Upon the increase in the magnetic flux, the vortices are still located near the regions with the suppressed diffusion coefficient whereas their cores are slightly shifted away from the axes of the cylinders due to the intervortex interaction (Fig.~\ref{Fig:cylinders_low_fields}(b)). Right at the first matching field $H = H^*$ the resulting vortex lattice perfectly matches the $D(\mathbf{r})$ distribution (Fig.~\ref{Fig:cylinders_low_fields}(c)). In the considered weak-field regime the vortex-occupation number for a particular defect $n_o$ (the number of vortices sitting on a defect) can take the following values: $n_o = 0, 1$.

\begin{figure*}[htpb]
\centering
\includegraphics[scale = 0.73]{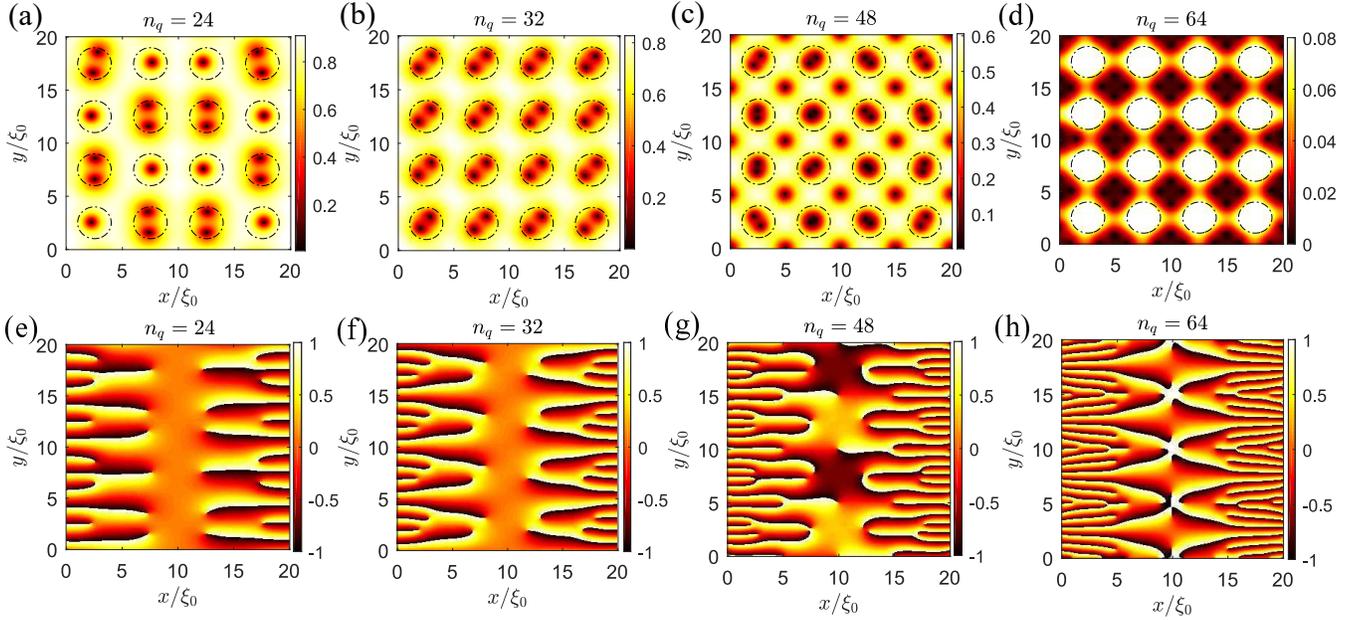}
\caption{Typical spatial profiles of the pair wave function for periodic $D(\mathbf{r})$ distribution shown in Fig.~\ref{Fig:diffusion_distribution}(a). (a-d) Color plots of $|\Psi(x,y)|$. (e-h) Color plots of the phase of the Cooper pair wave function $\chi(x,y)/\pi$. Panels (a, e), (b, f), (c, g), and (d, h) correspond to $n_q = 24$, $32$, $48$, and $64$, respectively. In panel (d) we have introduced the upper cutoff at $|\Psi| = 0.08$ for better visualization of the interstitial vortices. The parameters are: $T = 0.35T_{c0}$, $D_0/D_m = 3$, $R = 1.5\xi_0$, and $d = 2\xi_0$.}
\label{Fig:cylinders_high_fields}
\end{figure*}

Typical spatial profiles of the superconducting order parameter in the opposite case $|H|>H^*$ are presented in Fig.~\ref{Fig:cylinders_high_fields}. Panels (a), (b), (c), and (d) show the color plots of the absolute value of the pair wave function for $n_q = 24$, $32$, $48$, and $64$, respectively. Note that in panel (d) we have introduced the upper cutoff at $|\Psi| = 0.08$ for better visualization of the interstitial vortices. For clarity, we also reveal the color plots of the phase of the Cooper-pair wave function $\chi(x,y)/\pi$ in panels (e-h). In the case when the vortices are located in the regions with significantly suppressed Cooper pair density, their arrangement can be clearly seen on the spatial profiles of the phase $\chi(x,y)$. The obtained results show the way the vortex structure is modified upon further increase in the external magnetic field. In particular, for $n_q = 24$ (Figs.~\ref{Fig:cylinders_high_fields}(a) and~\ref{Fig:cylinders_high_fields}(e)), we see that the half of the defects now host two Abrikosov vortices while the vortex-occupation number for the rest of the defects $n_o = 1$. Right at the second matching field $n_q = 32$ (Figs.~\ref{Fig:cylinders_high_fields}(b) and~\ref{Fig:cylinders_high_fields}(f)) each region with the suppressed diffusion coefficient hosts a two-vortex molecule. One can see that for $n_q = 48$ (Figs.~\ref{Fig:cylinders_high_fields}(c) and~\ref{Fig:cylinders_high_fields}(g)) there also appear singly-quantized interstitial vortices which have the arrangement of a square lattice with the same lattice constant as for the $D(\mathbf{r})$ distribution. In a qualitative agreement with our previous results regarding the nucleation of superconductivity at isolated inhomogeneities (see Fig.~\ref{Fig:2D}), we observe that at sufficiently strong magnetic fields the vortex-occupation number for all of the defects becomes zero. Correspondingly, all of the vortices are located only between the defect regions. Typical spatial distribution of the superconducting order parameter for this case is shown in Figs.~\ref{Fig:cylinders_high_fields}(d) and~\ref{Fig:cylinders_high_fields}(h). One can clearly see that the interstitial vortices form a square lattice with four vortices per unit cell.

\begin{figure}[htpb]
\centering
\includegraphics[scale = 1.1]{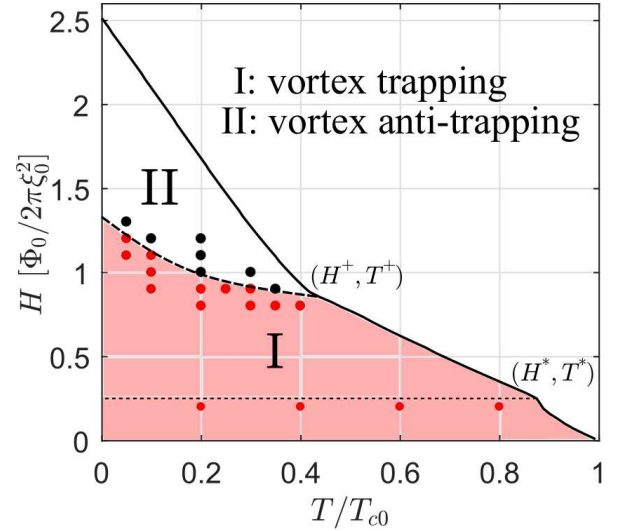}
\caption{Typical magnetic field - temperature phase diagram for superconductors with periodic $D(\mathbf{r})$ profile shown in Fig.~\ref{Fig:diffusion_distribution}(a). Solid line shows the temperature behavior of
the upper critical magnetic field $H_{c2}$. Dotted line shows the first matching field $H^*$. The parameters are: $D_0/D_m = 3$, $R = 1.5\xi_0$, and $d = 2\xi_0$. Filled circles represent the results of numerical simulations. In the region I (II) the regions with the suppressed diffusion coefficient attract (repel) vortices.}
\label{Fig:cylinders_phase_diagram}
\end{figure}

\begin{figure*}[htpb]
\centering
\includegraphics[scale = 0.84]{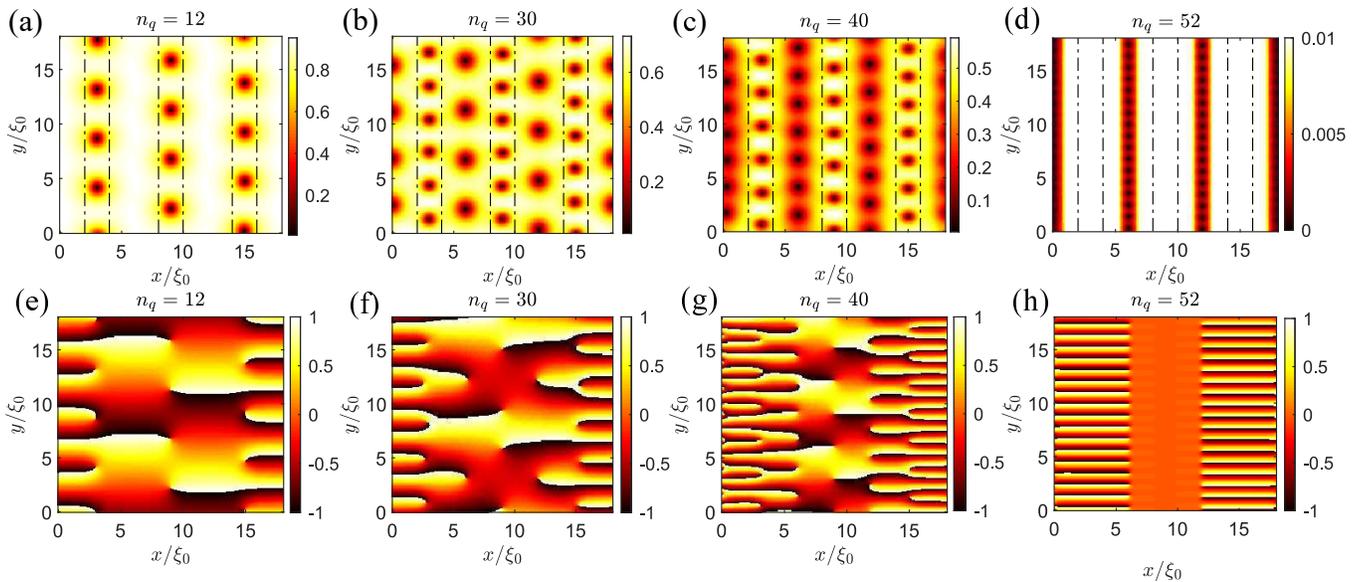}
\caption{Typical spatial profile of the Cooper-pair wave function for periodic $D(\mathbf{r})$ distribution shown in Fig.~\ref{Fig:diffusion_distribution}(b). (a-d) Color plots of $|\Psi(x,y)|$. (e-h) Color plots of the phase of the order parameter $\chi(x,y)/\pi$. Panels (a, e), (b, f), (c, g), and (d, h) correspond to $n_q = 12$, $30$, $40$, and $52$, respectively. In panel (d) we introduced the upper cutoff at $|\Psi| = 0.01$ for better visualization of the interstitial vortices. The parameters are: $T = 0.4T_{c0}$, $D_0/D_m = 3$, $\ell_c = 2\xi_0$, and $d = 4\xi_0$.}
\label{Fig:one_dimensional_order_parameter}
\end{figure*}

Our main findings in this subsection are summarized in Fig.~\ref{Fig:cylinders_phase_diagram}, where we plot the resulting phase diagram magnetic field $H$ - temperature $T$ for the considered set of the structure parameters $D_0/D_m = 3$, $R = 1.5\xi_0$, and $d = 2\xi_0$. Solid line shows the temperature behavior of the upper critical magnetic field $H_{c2}$. Dotted line shows the first matching field $H^*$. As a result, we find that the $H_{c2}(T)$ line possesses two features one of which is at the first matching field ($H^*,T^*$) and the other one is in the strong-field regime ($H^+,T^+$). Based on the results of numerical simulations, we identify two regions on the phase diagram with qualitatively different vortex arrangement (denoted as the regions I and II in Fig.~\ref{Fig:cylinders_phase_diagram}). Within the region I the defects attract Abrikosov vortices. Indeed, for rather weak magnetic fields $|H|<H^*$ all the vortices are pinned by the regions with the suppressed diffusion coefficient. In the opposite case $|H| > H^*$ (region I) the vortices are located at the defects and in between them. The switching between the vortex-defect attraction to the repulsion occurs at rather strong magnetic fields (region II). In this case all the vortices are located only between inhomogeneities (the vortices are pinned by the regions with larger diffusion coefficient). Note also that for the considered set of parameters, the values $H^+$ and $T^+$ for a superstructure deviate from the results for an isolated defect due to a rather small inter-defect distance $d$. In particular, we get $H^+/(\Phi_0/2\pi\xi_0^2) \approx 0.89$ and $T^+/T_{c0}\approx 0.42$ for a superstructure (see Fig.~\ref{Fig:cylinders_phase_diagram}) whereas the solution of the problem~(\ref{GL_linearized}) with $D(\mathbf{r})$ profile~(\ref{2D_diffusion_profile}) and $D_0/D_m = 3$, $R = 1.5\xi_0$ gives $H^+\approx 0.58$ and $T^+/T_{c0}\approx 0.54$.

\subsection{One-dimensional superconducting superlattice}

We continue with analyzing the vortex phases in one-dimensional superconducting superlattice subjected to the in-plane magnetic field (see Fig.~\ref{Fig:diffusion_distribution}(b)). The results of numerical simulations presented in this subsection have been obtained for a square computational cell covering three structure periods and the parameter set $D_0/D_m = 3$, $\ell_c = 2\xi_0$, and $d = 4\xi_0$.

\begin{figure}[htpb]
\centering
\includegraphics[scale = 1.1]{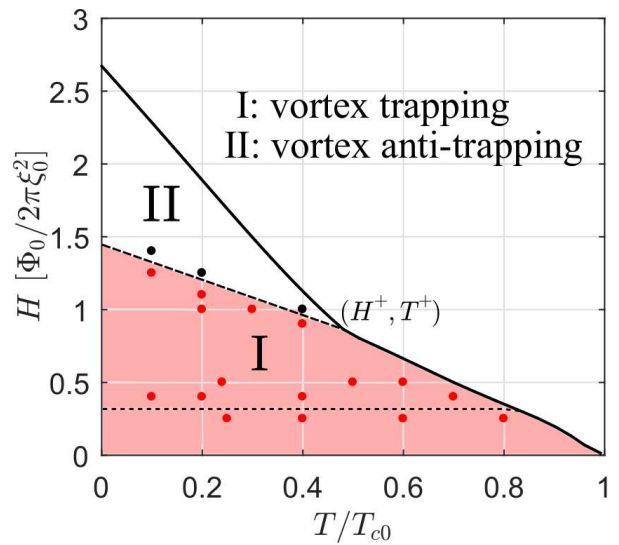}
\caption{Typical magnetic field - temperature phase diagram for the one-dimensional superconducting superlattice (see Fig.~\ref{Fig:diffusion_distribution}(b)). Solid line shows the temperature behavior of
the upper critical magnetic field $H_{c2}$. Dotted line shows the characteristic magnetic field, below which the vortices are located only in the regions with increased disorder. The parameters are: $D_0/D_m = 3$, $\ell_c = 2\xi_0$, and $d = 4\xi_0$. Filled circles represent the results of numerical simulations. In the region I (II) the defects attract (repel) Abrikosov vortices.}
\label{Fig:one_dimensional_phase_diagram}
\end{figure}

Typical spatial profiles of the superconducting order parameter for several values of the external magnetic fields are shown in Fig.~\ref{Fig:one_dimensional_order_parameter}. In particular, panels (a), (b), (c), and (d) reveal the color plots of the absolute values of the Cooper-pair wave function $|\Psi(x,y)|$ for $n_q = 12$, $30$, $40$, and $52$, respectively. Note that in panel (d) we have introduced the upper cutoff at $\Psi = 0.01$ for better visulatization of the interstitial vortices. Correponding color plots of the phase of the superconducting order parameter $\chi(x,y)/\pi$ are shown in panels (e-h). For rather weak applied magnetic fields $n_q = 12$ (Figs.~\ref{Fig:one_dimensional_order_parameter}(a) and~\ref{Fig:one_dimensional_order_parameter}(e)), we find that the vortices are pinned by the regions with the suppressed diffusion coefficient. The increase in $H$ leads to the rearrangement of the resulting vortex structure. One can see from Figs.~\ref{Fig:one_dimensional_order_parameter}(b) and~\ref{Fig:one_dimensional_order_parameter}(f) (see also Figs.~\ref{Fig:one_dimensional_order_parameter}(c) and~\ref{Fig:one_dimensional_order_parameter}(g)) that within the intermediate field range the vortex structure consists of two types of vortices with different core sizes located both in the regions with the suppressed diffusion constant and in between them. In a qualitative agreement with our previous results regarding the superconductivity nucleation at isolated defects (see Fig.~\ref{Fig:1D}(b)), we get that for rather strong magnetic fields the maximum pair density is reached in the layers with the increased impurity concentration whereas the layers with higher diffusion constant host chains of Abrikosov vortices (Figs.~\ref{Fig:one_dimensional_order_parameter}(d) and~\ref{Fig:one_dimensional_order_parameter}(h)).

Our main results of this subsection are presented in Fig.~\ref{Fig:one_dimensional_phase_diagram} where we plot a typical superconducting phase diagram magnetic field - temperature for the one-dimensional superlattice (see Fig.~\ref{Fig:diffusion_distribution}(b)). Solid line shows the temperature behavior of the parallel upper critical magnetic field $H_{c2||}(T)$ whereas the dotted line shows
the characteristic magnetic field, below which the vortices are located only in the regions with increased disorder. It is interesting to note that the resulting phase diagrams of one-dimensional superlattice possess some similarities with previously considered case of regular arrays of cylindrical defects or discs. Indeed, the resulting phase diagrams in Fig.~\ref{Fig:one_dimensional_phase_diagram} can be divided into two regions with qualitatively different vortex arrangement. Within the region I the defects attract Abrikosov vortices. In particular, for rather weak magnetic fields the vortices are located only inside the regions with increased disorder. Within the intermediate field range (region I) the vortex structure is formed by two types of vortices with different core sizes located both in the regions with the suppressed diffusion coefficient and in between them. Finally, at rather strong magnetic fields (region II) vortices are pinned by the regions with larger diffusion coefficient. Note also that for the chosen parameter set, the values $H^+$ and $T^+$ for the one-dimensional superlattice appear to be rather close to the results for an isolated defect due to a rather large interlayer distance $d$. In particular, we get $H^+/(\Phi_0/2\pi\xi_0^2) \approx 0.89$ and $T^+/T_{c0}\approx 0.42$ for a superstructure (see Fig.~\ref{Fig:one_dimensional_phase_diagram}) whereas the solution of the problem~(\ref{GL_linearized}) with the $D(\mathbf{r})$ profile~(\ref{1D_diffusion_profile}) and $D_0/D_m = 3$, $\ell_c = 2\xi_0$ gives $H^+/(\Phi_0/2\pi\xi_0^2)\approx 0.84$ and $T^+/T_{c0}\approx 0.48$.

\section{Concluding remarks}\label{discussion}

Experimentally, the phenomenon of switching from the vortex-defect attraction to the repulsion described in our work and the corresponding rearrangement of the vortex lattice can be directly verified using common vortex imaging techniques, which include the Bitter decoration technique~\cite{BezryadinPRB1996,HerringPL1974,BezryadinJLTP1996}, electron holography~\cite{BonevichPRL1993}, scanning probe microscopy~\cite{KarapetrovPRL2005,BehlerPRL1994,GrigorenkoPRB2001,FieldPRL2002,SilhanekPRB2011}, Lorentz microscopy~\cite{HaradaS1996}, and magneto-optical imaging~\cite{YurchenkoPC2006}. On the other hand, it is well known that the pinning properties of a particular structure can reveal themselves in features of the critical current density as a function of the applied magnetic field $J_c(H)$ (see, e.g., Refs.~\cite{BerdiyorovEPL2006,BerdiyorovPRB2006,SabatinoJAP2010,LatimerPRB2012,LatimerPRL2013,XueNJP2018,SadovskyyPRB2017,MetlushkoPRB21999,VanLookPRB2002}). In the context of our results obtained for one-dimensional superlattices, it is interesting to note that experimental measurements of the critical current density versus the magnetic field in Nb/NbZr multilayers~\cite{KoorevaarPRB1993,NojimaPB1994} identified two temperature-dependent lines $H_P(T)$ and $H_L(T)$ on the $H$-$T$ phase diagram of such devices (see Fig.~1 in Ref.~\cite{NojimaPB1994}). In the regime when both the applied current and the magnetic field lie in the plane of the layers and perpendicular to each other, the resulting $J_c(H)$ dependence has a peak at the $H_P(T)$ line while a change in the slope of the $J_c(H)$ dependence occurs at the $H_L(T)$ line. Presumably, the appearance of the peak on $J_c(H)$ curve at weak magnetic fields can be related to the well-known commensurability effects (see, e.g., Refs.~\cite{BerdiyorovEPL2006,BerdiyorovPRL2006,BerdiyorovPRB2006,SabatinoJAP2010,LatimerPRB2012,LatimerPRL2013,GePRB2017,XueNJP2018,BerdiyorovNJP2009,SadovskyyPRB2017,FieldPRL2002,SilhanekPRB2011,HaradaS1996,YurchenkoPC2006,MetlushkoPRB1999,MetlushkoPRB21999,VanLookPRB2002,BaertPRL1995,MoshchalkovPRB1996,RaedtsPRB2004,SilhanekPRB2004,RaedtsPRB2006,Cuadra-SolisPRB2014}), which manifest themselves through the enhancement of the pinning properties of the sample when the vortex lattice matches the defect array (vortex density and the defect density are integer multiples of each
other). On the other hand, it is tempting to associate the change in the slope of $J_c(H)$ curve at rather strong magnetic fields with the switching from the vortex-defect attraction to the repulsion. However, this detailed analysis of this question requires quantitative studies of the magnetotransport properties of superconductors with modulated disorder, which is behind the scope of the present work.

To sum up, we have uncovered and explained the phenomenon of switching between the vortex-defect attraction to the repulsion in
superconductors with modulated disorder. It has been shown that a superconducting nucleus localized near the region with the suppressed diffusion coefficient can possess a nonzero vorticiy whereas the increase in the applied magnetic field can result in the transition into the state with zero winding number. We have demonstrated the appearance of this switching phenomenon in superconductors with several periodic spatial profiles of the diffusion coefficient. Our results clarify the vortex arrangement in several classes of the superconducting materials including one-dimensional superlattices and nanopatterned superconductors with regular arrays of the defects characterized by the increased concentration of nonmagnetic impurities.

\acknowledgements
We thank V.~L. Vadimov, A.~A. Bespalov, I.~M.~Khaymovich, A.~V.~Samokhvalov and I.~A. Shereshevskii for stimulating discussions. This work was supported by the Russian Science Foundation (Grant No. 21-12-00409).

\appendix
\section{Details of numerical calculations of the phase-transition lines}\label{appendix}

Here we provide the details of numerical calculations of the phase-transition lines magnetic field - temperature for deterministic $D(\mathbf{r})$ profiles~(\ref{diffusion_profiles}). Considering the two-dimensional profile~(\ref{2D_diffusion_profile}) and choosing the radial gauge $\mathbf{A} = H\rho\boldsymbol{\varphi}_0/2$, we seek the solution of Eq.~(\ref{GL_linearized}) in the form
\begin{equation}\label{2D_order_parameter}
 \Psi(\mathbf{r}) = \psi_{n}(\rho)e^{in\varphi} \ .
\end{equation}
Substitution of Eq.~(\ref{2D_order_parameter}) into Eq.~(\ref{GL_linearized}) yields the following equations for the radial part of the Cooper pair wave function in the regions with a constant diffusion coefficient:
\begin{subequations}\label{2D_equation}
\begin{align}
 H_n(\rho)\psi_n(\rho) = [E_0/\xi^2(\rho)]\psi_n(\rho) \ ,\\
 H_n(\rho) = \biggl[-\frac{1}{\rho}\frac{d}{d\rho}\left(\rho\frac{d}{d\rho}\right) 
 + \frac{1}{\rho^2}\left(n + \frac{\rho^2}{2L_H^2}\right)^2\biggl] \ .
 \end{align}
\end{subequations}
Here $\xi(\rho) = \xi_0$ ($\xi_m$)  for $\rho > R$ ($\rho < R$), and $\xi_{0,m} = \sqrt{\pi\hbar D_{0,m}/8T_{c0}}$. The solutions of Eq.~(\ref{2D_equation}) read
\begin{subequations}\label{solutions_2D_profile}
\begin{align}
 \psi_{n}^I = C_1 e^{-\rho^2/4 L_H^2}\rho^{|n|} L^{|n|}_{a_m}\left(\frac{\rho^2}{2 L_H^2}\right) \ ,\\
 \psi_{n}^{II} = C_2 e^{-\rho^2/4 L_H^2}\rho^{|n|} U\left(-{a_0}, |n| + 1, \frac{\rho^2}{2 L_H^2}\right) \ ,
 \end{align}
\end{subequations}
where the region I (II) is defined by $\rho<R$ ($\rho > R$), $L^a_n(x)$ is the generalized Laguerre polynomial, $U(a,b,x)$ is the Tricomi's confluent hypergeometric function, and
\begin{equation}\label{function_arguments}
 a_{0,m}(n) = \frac{E_0L_H^2}{2\xi_{0,m}^2} - \frac{1 + n + |n|}{2} \ .
\end{equation}
Imposing the boundary conditions 
\begin{subequations}
\begin{align}
    \psi_n^I(R) = \psi_n^{II}(R) \ , \\ 
    D_m \frac{d\psi_n^I}{d\rho}\biggl|_{\rho = R} = D_0 \frac{d\psi_n^{II}}{d\rho}\biggl|_{\rho = R} \ ,
  \end{align}
\end{subequations}
on the solutions~(\ref{solutions_2D_profile}), we get a homogeneous system of linear equations for the coefficients $C_1$ and $C_2$. Putting the determinant of the resulting system to be zero, we find the lowest eigenvalue of the problem~(\ref{GL_linearized}).

Let us now consider the one-dimensional profile of the diffusion coefficient~(\ref{1D_diffusion_profile}). Choosing the Landau gauge $\mathbf{A} = (0,Hx,0)$ and substituting the Cooper-pair wave function 
\begin{equation}
 \Psi(\mathbf{r}) = \psi_{k_y}(x)e^{ik_yy} \ ,
\end{equation}
into Eq.~(\ref{GL_linearized}), we get the following equations in the regions with a constant diffusion coefficient:
\begin{equation}\label{1D_equation}
 \left(-\frac{d^2}{dx^2} + \frac{\tilde{x}^2}{L_H^4}\right)\psi_{k_y}(x) = \frac{E_0}{\xi^2(x)}\psi_{k_y}(x) \ .
\end{equation}
Here $\tilde{x} = x - x_0$, $x_0 = -k_yL_H^2$, and $\xi(x) = \xi_m$ ($\xi_0$) for $|x|<\ell_c/2$ ($|x|>\ell_c/2$). The solutions of the above equation can be written as follows
\begin{subequations}\label{solutions_1D_profile}
 \begin{align}
  \psi_{k_y}^I = C_1 e^{-\tilde{x}^2/2L_H^2}H_{n_0}\left(-\frac{\tilde{x}}{L_H}\right) \ ,\\
  \psi_{k_y}^{II} = C_2 e^{-\tilde{x}^2/2L_H^2} {_1}F_1\left[-\frac{n_m}{2}, \frac{1}{2}, \left(\frac{\tilde{x}}{L_H}\right)^2\right]\\
  \nonumber
  +C_3 e^{-\tilde{x}^2/2L_H^2} x  _1F_1\left[-\frac{n_m-1}{2},\frac{3}{2}, \left(\frac{\tilde{x}}{L_H}\right)^2\right] \ ,\\
  \psi_{k_y}^{III} = C_4 e^{-\tilde{x}^2/2L_H^2}H_{n_0}\left(\frac{\tilde{x}}{L_H}\right) \ ,
 \end{align} 
\end{subequations}
where the regions I, II, and III correspond to $x<-\ell_c/2$, $|x|<\ell_c/2$, and $x>\ell_c/2$, respectively, ${_1}F_1(a,b,x)$ is the confluent hypergeometric function, $H_{\nu}(x)$ is the Hermite function of degree $\nu$~\cite{LebedevBook}. The quantities $n_{0,m}$ in Eqs.~(\ref{solutions_1D_profile}) are defined by Eq.~(\ref{function_arguments}) for $n = 0$. Imposing the boundary conditions
\begin{subequations}
 \begin{align}
  \psi^I_{k_y}(-\ell_c/2) = \psi^{II}_{k_y}(-\ell_c/2) \ ,\\
  \psi^{II}_{k_y}(\ell_c/2) = \psi^{III}_{k_y}(\ell_c/2) \ ,\\
  D_0\frac{d\psi^I_{k_y}}{dx}\biggl|_{x = -\ell_c/2} = D_m\frac{d\psi^{II}_{k_y}}{dx}\biggl|_{x = -\ell_c/2} \ ,\\
  D_m\frac{d\psi^{II}}{dx}\biggl|_{x = \ell_c/2} = D_0\frac{d\psi^{III}}{dx}\biggl|_{x = \ell_c/2} \ ,
 \end{align}
\end{subequations}
on the solutions~(\ref{solutions_1D_profile}), we get a homogeneous system of linear equations for the coefficients $C_1$, $C_2$, $C_3$, and $C_4$. Putting the determinant of the resulting system to be zero, we determine the lowest eigenvalue of the problem~(\ref{GL_linearized}).


\begin{references}
  \bibitem{Campbell1972} A.~Campbell and J.~E.~Evetts, \textit{Critical Currents in Superconductors} (Taylor and Francis, London, 1972).
  \bibitem{Blatter1994} G.~Blatter, M.~V.~Feigel'man, V.~B.~Geshkenbein, A.~I.~Larkin, and V.~M.~Vinokur, Vortices in high-temperature superconductors, Rev. Mod. Phys. \textbf{66}, 1125 (1994).
  
  \bibitem{BezryadinPLA1994} A.~Bezryadin, A.~Buzdin, B.~Pannetier, Phase transitions in a superconducting thin film with a single circular hole, Phys. Lett. A \textbf{195}, 373 (1994).
  \bibitem{BezryadinJLTP1995} A.~Bezryadin and B.~Pannetier, Nucleation of Superconductivity in a Thin Film with a Lattice of Circular Holes, J. Low Temp. Phys. \textbf{98}, 251 (1995).
  \bibitem{BerdiyorovEPL2006} G.~R.~Berdiyorov, M.~V.~Milo\v{s}evi\'{c}, and F.~M.~Peeters, Superconducting films with antidot arrays - Novel behavior of the critical current, Europhys. Lett. \textbf{74}, 493 (2006).
  \bibitem{BerdiyorovPRL2006} G.~R.~Berdiyorov, M.~V.~Milo\v{s}evi\'{c}, and F.~M.~Peeters, Novel Commensurability Effects in Superconducting Films with Antidot Arrays, Phys. Rev. Lett. \textbf{96}, 207001 (2006).
  \bibitem{BerdiyorovPRB2006} G.~R.~Berdiyorov, M.~V.~Milo\v{s}evi\'{c}, and F.~M.~Peeters, Vortex configurations and critical parameters in superconducting thin films containing antidot arrays: Nonlinear Ginzburg-Landau theory, Phys. Rev. B \textbf{74}, 174512 (2006).
  \bibitem{SabatinoJAP2010} P.~Sabatino, C.~Cirillo, G.~Carapella, M.~Trezza, and C.~Attanasio, High field vortex matching effects in superconducting Nb thin films with a nanometer-sized square array of antidots, J. Appl. Phys. \textbf{108}, 053906 (2010).
  \bibitem{LatimerPRB2012} M.~L.~Latimer, G.~R.~Berdiyorov, Z.~L.~Xiao, W.~K.~Kwok, and F.~M.~Peeters, Vortex interaction enhanced saturation number and caging effect in a superconducting film with a honeycomb array of nanoscale holes, Phys. Rev. B \textbf{85}, 012505 (2012).
  \bibitem{LatimerPRL2013} M.~L.~Latimer, G.~R.~Berdiyorov, Z.~L.~Xiao, F.~M.~Peeters, and W.~K.~Kwok, Realization of Artificial Ice Systems for Magnetic Vortices in a Superconducting MoGe Thin Film with Patterned Nanostructures, Phys. Rev. Lett. \textbf{111}, 067001 (2013).
  \bibitem{GePRB2017} J.-Y.~Ge, V.~N~Gladilin, J.~Tempere, V.~S.~Zharinov, J.~Van de Vondel, J.~T.~Devreese, and V.~V.~Moshchalkov, Direct visualization of vortex ice in a nanostructured superconductor, Phys. Rev. B \textbf{96}, 134515 (2017).
  \bibitem{XueNJP2018} C.~Xue, J.-Y.~Ge, A.~He, V.~S.~Zharinov, V.~V.~Moshchalkov, and Y.-H.~Zhou, Stability of degenerate vortex states and multi-quanta confinement effects in a nanostructured superconductor with Kagome lattice of elongated antidots, New J. Phys. \textbf{20}, 093030 (2018).
  \bibitem{BezryadinPRB1996} A.~Bezryadin, Yu.~N.~Ovchinnikov, and B.~Pannetier, Nucleation of vortices inside open and blind microholes, Phys. Rev. B \textbf{53}, 8553 (1996).
  \bibitem{BerdiyorovNJP2009} G.~R.~Berdiyorov, M.~V.~Milo\v{s}evi\'{c}, and F.~M.~Peeters, Composite vortex ordering in superconducting films with arrays of blind holes, New J. Phys. \textbf{11}, 013025 (2009).
  \bibitem{KarapetrovPRL2005} G.~Karapetrov, J.~Fedor, M.~Iavarone, D.~Rosenmann, and W.~K.~Kwok, Direct Observation of Geometrical Phase Transitions in Mesoscopic Superconductors by Scanning Tunneling Microscopy, Phys. Rev. Lett. \textbf{95}, 167002 (2005).
  \bibitem{SadovskyyPRB2017} I.~A.~Sadovskyy, Y.~L.~Wang, Z.-L.~Xiao, W.-K.~Kwok, and A.~Glatz, Effect of hexagonal patterned arrays and defect geometry on the critical current of superconducting films, Phys. Rev. B \textbf{95}, 075303 (2017).
 
  \bibitem{BeanPRL1971} C.~P.~Bean and J.~D.~Livingston, Surface Barrier in Type-II Superconductors, Phys. Rev. Lett. \textbf{12}, 14 (1964).
 
  \bibitem{BugoslavskyN2001} Y.~Bugoslavsky, L.~F.~Cohen, G.~K.~Perkins, M.~Polichetti, T.~J.~Tate, R.~Gwilliam, and A.~D.~Caplin, Enhancement of the high-magnetic-field critical density of superconducting MgB$_2$ by proton irradiation, Nature \textbf{411}, 561 (2001).
  \bibitem{NakajimaPRB2009} Y.~Nakajima, Y.~Tsuchiya, T.~Taen, T.~Tamegai, S.~Okayasu, and M.~Sasase, Enhancement of critical current density in Co-doped BaFe$_2$As$_2$ with columnar defects introduced by heavy-ion irradiation, Phys. Rev. B \textbf{80}, 012510 (2009).
  \bibitem{ZechnerSUST2018} G.~Zechner, K.~L.~Mletschnig, W.~Lang, M.~Dosmailov, M.~A.~Bodea, and J.~D.~Pedarnig, Unconventional critical state in YBa$_{2}$Cu$_{3}$O$_{7-\delta}$ thin films with a vortex-pin lattice fabricated by masked He$^+$ ion beam irradiation, Supercond. Sci. Technol. \textbf{31}, 044002 (2018).
  \bibitem{AntonovPSS2019} A.~V.~Antonov, A.~V.~Ikonnikov, D.~V.~Masterov, A.~N.~Mikhaylov, S.~V.~Morozov, Yu.~N.~Nozdrin, S.~A.~Pavlov, A.~E.~Parafin, D.~I.~Tetel'baum, S.~S.~Ustavschikov, P.~A.~Yunin, and D.~A.~Savinov, Phase Diagrams of Thin Disordered Films Based on HTSC YBa$_{2}$Cu$_{3}$O$_{7-x}$ in External Magnetic Fields, Phys. Solid State \textbf{61}, 1523 (2019).
  \bibitem{AntonovPSS2020} A.~V.~Antonov, A.~I.~El'kina, V.~K.~Vasiliev, M.~A.~Galin, D.~V.~Masterov, A.~N.~Mikhaylov, S.~V.~Morozov, S.~A.~Pavlov, A.~E.~Parafin, D.~I.~Tetelbaum, S.~S.~Ustavschikov, P.~A.~Yunin, and D.~A.~Savinov, Experimental Observation of s-Component of Superconducting Pairing in Thin Disordered HTSC Films Based on YBCO, Phys. Solid State \textbf{62}, 1598 (2020).
  \bibitem{AntonovPC2020} A.~V.~Antonov, A.~V.~Ikonnikov, D.~V.~Masterov, A.~N.~Mikhaylov, S.~V.~Morozov, Yu.~N.~Nozdrin, S.~A.~Pavlov, A.~E.~Parafin, D.~I.~Tetel'baum, S.~S.~Ustavschikov, V.~K.~Vasiliev, P.~A.~Yunin, D.~A.~Savinov, Critical-field slope reduction and upward curvature of the phase-transition lines of thin disordered superconducting YBa$_2$Cu$_{3}$O$_{7-x}$ films in strong magnetic fields, Phys. C: Supercond. Appl. \textbf{568}, 1353581 (2020).
  \bibitem{AichnerFNT2020} B.~Aichner, K.~L.~Mletschnig, B.~M\"{u}ller, M.~Karrer, M.~Dosmailov, J.~D.~Pedarnig, R.~Kleiner, D.~Koelle, and W.~Lang, Angular magnetic-field dependence of vortex matching in pinning lattices fabricated by focused or masked helium ion beam irradiation of superconducting YBa$_2$Cu$_3$O$_{7-\delta}$ thin films, Fiz. Nizk. Temp. \textbf{46}, 402 (2020).

  \bibitem{ThunebergPRB1984} E.~V.~Thuneberg, J.~Kurij\"{a}rvi, and D.~Rainer, Elementary-flux-pinning potential in type-II superconductors, Phys. Rev. B \textbf{29}, 3913 (1984).
 
 
  \bibitem{TakahashiPRB11986} S.~Takahashi and M.~Tachiki, Theory of the upper critical field of superconducting superlattices, Phys. Rev. B \textbf{33}, 4620 (1986).
  \bibitem{TakahashiPRB21986} S.~Takahashi and M.~Tachiki, New phase diagram in superconducting superlattices, Phys. Rev. B \textbf{34}, 3162 (1986).
  \bibitem{KopasovRQE2017} A.~A.~Kopasov, D.~A.~Savinov, and A.~S.~Mel'nikov, Localized Superconductivity in Systems with Inhomogeneous Mass of Cooper Pairs, Radiophys. Quantum Electron. \textbf{59}, 911 (2017).
  \bibitem{KopasovPRB2017} A.~A.~Kopasov, D.~A.~Savinov, and A.~S.~Mel'nikov, Crossover between Abrikosov vortex lattice and superconducting droplet state in superconductors with modulated disorder, Phys. Rev. B \textbf{95}, 104520 (2017).
  
  \bibitem{LarkinJETP1970} A.~I.~Larkin, Effect of inhomogeneities on the structure of the mixed state of superconductors, Sov. Phys. JETP \textbf{31}, 784 (1970).
  \bibitem{LarkinJETP1972} A.~I.~Larkin and Yu.~N.~Ovchinnikov, Influence of inhomogeneities on superconducting properties, Sov. Phys. JETP \textbf{34}, 651 (1972).
  
  \bibitem{Anderson1959} P.~W.~Anderson, Theory of dirty superconductors, J. Phys. Chem. Solids \textbf{11}, 26 (1959).
  \bibitem{Abrikosov1959} A.~A.~Abrikosov and L.~P.~Gor'kov, Superconducting alloys at finite temperatures, Zh. Eksp. Teor. Fiz. \textbf{36}, 319 (1959) [Sov. Phys. JETP \textbf{9}, 220 (1959)].
  
  \bibitem{AmiPTP1975} S.~Ami and K.~Maki, Pinning Effect due to Periodic Variation of Impurity Concentration in Type II Superconductors, Prog. Theor. Phys. \textbf{53}, 1 (1975).
  \bibitem{YuanZP1995} B.~J.~Yuan, Mixed state in a superlattice, Z. Phys. \textbf{98}, 457 (1995).
  
  \bibitem{KarkutPRL1988} M.~G.~Karkut, V.~Matijasevic, L.~Antognazza, J.-M.~Triscone, N.~Missert, M.~R.~Beasley, and \O{}.~Fischer, Anomalous upper critical fields of superconducting multilayers: Verification of the Takahashi-Tachiki effect, Phys. Rev. Lett. \textbf{60}, 1751 (1988).
  \bibitem{KuwasawaPC1990} Y.~Kuwasawa, U.~Hayano, T.~Tosaka, S.~Nakano, and S.~Matuda, Observation of anomalous transition in the upper critical fields of Nb/Nb$_{0.5}$Zr$_{0.5}$ multilayers, Phys. C: Supercond. \textbf{165}, 173 (1990).
  \bibitem{AartsPB1990} J.~Aarts, K.-J.~de Korver, W.~Maj, and P.~H.~Kes, Parallel critical fields in Nb/Nb$_{0.6}$Zr$_{0.4}$ multilayers, Phys. B: Condens. Matter \textbf{165}, 475 (1990).
  \bibitem{AartsEPL1990} J.~Aarts, K.J.~de Korver, and P.~H.~Kes, Dimensionality Crossovers in the Parallel Critical Fields of Nb/Nb$_{0.6}$Zr$_{0.4}$ Multilayers, Europhys. Lett. \textbf{12}, 447 (1990).
  \bibitem{KuwasawaPC1991} Y.~Kuwasawa, T.~Tosaka, A.~Uchiyama, S.~Matuda, and S.~Nakano, Anisotropy of critical current and anomalous upturn feature of parallel upper critical fields in the superconducting Nb/Nb$_{0.5}$Zr$_{0.5}$ multilayer, Phys. C: Supercond. \textbf{175}, 187 (1991).
  \bibitem{NojimaPC1993} T.~Nojima, M.~Kinoshita, S.~Nakano, and Y.~Kuwasawa, Transport critical current density and dimensional crossover in superconducting Nb/NbZr multilayers, Phys. C: Supercond. \textbf{206}, 387 (1993).
  \bibitem{KoorevaarPRB1993} P.~Koorevaar, W.~Maj, P.~H.~Kes, and J.~Aarts, Vortex-lattice transition in superconducting Nb/NbZr multilayers, Phys. Rev. B \textbf{47}, 934 (1993).
  \bibitem{NojimaPB1994} T.~Nojima, M.~Kinoshita, S.~Nakano, and Y.~Kuwasawa, Vortex-Lattice Phase Diagram of Nb/NbZr Multilayers, Phys. B: Condens. Matter \textbf{194}, 1879 (1994).
  \bibitem{KuwasawaPB1996} Y.~Kuwasawa, T.~Nojima, S.~Hwang, B.~J.~Yuan, J.~P.~Whitehead, Preferential position for nucleation of order parameter and angular dependence of upper critical fields in Nb/NbZr multilayers, Phys. B: Condens. Matter \textbf{222}, 92 (1996).
  \bibitem{ObiPSS2001} Y.~Obi, M.~Ikebe, H.~Fijishiro, K.~Takanaka, and H.~Fujimori, Synthetic Analyses of T$_c$ and H$_{c2}$ of NbTi/Nb Superconductor/Superconductor Superlattice, Phys. Stat. Sol. \textbf{223}, 799 (2001).
  
  \bibitem{SadovskiiJCP2015} I.~A.~Sadovskii, A.~E.~Koshelev, C.~L.~Phillips, D.~A.~Karpeyev, and A.~Glatz, Stable large-scale solver for Ginzburg-Landau equations for superconductors, J. Comp. Phys. \textbf{294}, 639 (2015).
  
  \bibitem{KatoPRB1993} R.~Kato, Y.~Enomoto, and S.~Maekawa, Effects of the surface boundary on the magnetization process in type-II superconductors, Phys. Rev. B \textbf{47}, 8016 (1993).
  
  \bibitem{SJ} D.~Saint-James, G.~Sarma, and E.~J.~Thomas, \textit{Type II superconductivity} (Oxford Pergamon Press, 1969).
  
  \bibitem{HerringPL1974} C.~P.~Herring, The observation of flux line pinning in superconducting foils, Phys. Lett. \textbf{47A}, 105 (1974).
  \bibitem{BezryadinJLTP1996} A.~Bezryadin and B.~Pannetier, Role of edge superconducting states in trapping of multi-quanta vortices by microholes. Application of the bitter decoration technique, J. Low Temp. Phys. \textbf{102}, 73 (1996).
  \bibitem{BonevichPRL1993} J.~E.~Bonevich, K.~Harada, T.~Matsuda, H.~Kasai, T.~Yoshida, G.~Pozzi, and A.~Tonomura, Electron holography observation of vortex lattices in a superconductor, Phys. Rev. Lett. \textbf{70}, 2952 (1993).
  \bibitem{BehlerPRL1994} S.~Behler, S.~H.~Pan, P.~Jess, A.~Baratoff, H.~J.~Guntherodt, F.~Levy, G.~Wirth, and J.~Wiesner, Vortex Pinning in Ion-Irradiated NbSe$_2$ Studied by Scanning Tunneling Microscopy, Phys. Rev. Lett. \textbf{72}, 1750 (1994).
  \bibitem{GrigorenkoPRB2001} A.~N.~Grigorenko, G.~D.~Howells, S.~J.~Bending, J.~Bekaert, M.~J.~Van Bael, L.~Van Look, V.~V.~Moshchalkov, Y.~Bruynseraede, G.~Borghs, I.~I.~Kaya, and R.~A.~Stradling, Direct imaging of commensurate vortex structures in ordered antidot arrays, Phys. Rev. B \textbf{63}, 052504 (2001).
  \bibitem{FieldPRL2002} S.~B.~Field, S.~S.~James, J.~Barentine, V.~Metlushko, G.~Crabtree, H.~Shtrikman, B.~Ilic, and S.~R.~J.~Brueck, Vortex Configurations, Matching, and Domain Structure in Large Arrays of Artificial Pinning Centers, Phys. Rev. Lett. \textbf{88}, 067003 (2002).
  \bibitem{SilhanekPRB2011} A.~V.~Silhanek, J.~Gutierrez, R.~B.~G.~Kramer, G.~W.~Ataklti, J.~Van de Vondel, V.~V.~Moshcalkov, and A.~Sanchez, Microscopic picture of the critical state in a superconductor with a periodic array of antidots, Phys. Rev. B \textbf{83}, 024509 (2011).
  \bibitem{HaradaS1996} K.~Harada, O.~Kamimura, H.~Kasai, T.~Matsuda, A.~Tonomura, and V.~V.~Moshchalkov, Direct Observation of Vortex Dynamics in Superconducting Films with Regular Arrays of Defects, Science \textbf{274}, 1167 (1996).
 
  \bibitem{YurchenkoPC2006} V.~V.~Yurchenko, R.~W\"{o}rdenweber, Yu.~M.~Galperin, D.~V.~Shantsev, J.~I.~Vestg\r{a}rden, and T.~H.~Johansen, Magneto-optical imaging of magnetic flux patterns in superconducting films with antidots, Physica C \textbf{437}, 357 (2006).

 
  \bibitem{MetlushkoPRB1999} V.~Metlushko, U.~Welp, G.~W.~Crabtree, Z.~Zhang, S.~R.~J.~Brueck, B.~Watkins, L.~E.~DeLong, B.~Ilic, K.~Chung, and P.~J.~Hesketh, Nonlinear flux-line dynamics in vanadium films with square lattices of submicron holes, Phys. Rev. B \textbf{59}, 603 (1999).
  \bibitem{MetlushkoPRB21999} V.~Metlushko, U.~Welp, G.~W.~Crabtree, R.~Osgood, S.~D.~Bader, L.~E.~DeLong, Z.~Zhang, S.~R.~J.~Brueck, B.~Ilic, K.~Chung, and P.~J.~Hesketh, Interstitial flux phases in a superconducting niobium film with a square lattice of artificial pinning centers, Phys. Rev. B \textbf{60}, 12585(R) (1999).
  \bibitem{VanLookPRB2002} L.~Van Look, B.~Y.~Zhu, R.~Jonckheere, B.~R.~Zhao, Z.~X.~Zhao, and V.~V.~Moshchalkov, Anisotropic vortex pinning in superconductors with a square array of rectangular submicron holes, Phys. Rev. B \textbf{66}, 214511 (2002).
  \bibitem{BaertPRL1995} M.~Baert, V.~V.~Metlushko, R.~Jonkheere, V.~V.~Moshchalkov, and Y.~Bruynseraede, Composite Flux-Line Lattices Stabilized in Superconducting Films by a Regular Array of Artificial Defects, Phys. Rev. Lett. \textbf{74}, 3269 (1995).
  \bibitem{MoshchalkovPRB1996} V.~V.~Moshchalkov, M.~Baert, V.~V.~Metlushko, E.~Rosseel, M.~J.~Van Bael, K.~Temst, R.~Jonckheere, and Y.~Bruynseraede, Magnetization of multiple-quanta vortex lattices, Phys. Rev. B \textbf{54}, 7385 (1996).  
  \bibitem{RaedtsPRB2004} S.~Raedts, A.~V.~Silhanek, M.~J.~Van Bael, and V.~V.~Moshchalkov, Flux-pinning properties of superconducting films with arrays of blind holes, Phys. Rev. B \textbf{70}, 024509 (2004).
  \bibitem{SilhanekPRB2004} A.~V.~Silhanek, S.~Raedts, M.~J.~Van Bael, and V.~V.~Moshchalkov, Experimental determination of the number of flux lines trapped by microholes in superconducting samples, Phys. Rev. B \textbf{70}, 054515 (2004).
  \bibitem{RaedtsPRB2006} S.~Raedts, A.~V.~Silhanek, V.~V.~Moshchalkov, J.~Moonens, and L.~H.~A.~Leunissen, Crossover from intravalley to intervalley vortex motion in type-II superconductors with a periodic pinning array, Phys. Rev. B \textbf{73}, 174514 (2006).
  \bibitem{Cuadra-SolisPRB2014} P.-d.-J.~Cuadra-Sol\'{i}s, A.~Garc\'{i}a-Santiago, J.~M.~Hernandez, J.~Tejada, J.~Vanacken, and V.~V.~Moshchalkov, Observation of commensurability effects in a patterned thin superconducting Pb film using microwave reflection spectrometry, Phys. Rev. B \textbf{89}, 054517 (2014).


  \bibitem{LebedevBook} N.~N.~Lebedev, \textit{Special Functions and Their Applications} (New York: Dover publications, 1972).


 \end{references}
\end{document}